\documentclass[12pt]{article}
\usepackage[latin9]{inputenc}
\usepackage{amsmath}
\usepackage{amssymb,color}
\usepackage{graphicx}
\usepackage{esint}

\makeatletter
\usepackage{subfigure}\usepackage{amscd}
\usepackage{appendix}
\usepackage{verbatim}
\usepackage{epstopdf}

\usepackage{color}

\makeatother

\begin{document}
\begin{titlepage}
OCHA-PP-351 \hspace{-3.5cm}{\hbox to\hsize{\hfill{}July 2018 }}

\bigskip{}
\vspace{3\baselineskip}

\begin{center}
\textbf{\Large{Pendulum Leptogenesis}}\par
\end{center}{\Large \par}

\begin{center}
\textbf{ Kazuharu Bamba${}^{1*}$,
	Neil D. Barrie${}^{2\dagger}$,
	Akio Sugamoto${}^{3,4\ddagger}$,\\
	Tatsu Takeuchi${}^{5\S}$, 
	Kimiko Yamashita${}^{3,6,7,8\P}$
 }\\
\textbf{ }
\par\end{center}

\begin{center}
{\it
	${}^{1}$Division of Human Support System, Faculty of Symbiotic Systems Science,\\
	Fukushima University, Fukushima 960-1296, Japan\\
	${}^{2}$Kavli IPMU (WPI), UTIAS, University of Tokyo, Kashiwa, Chiba 277-8583, Japan\\
	${}^{3}$Department of Physics, Graduate School of Humanities and Sciences,\\
	Ochanomizu University, 2-1-1 Ohtsuka, Bunkyo-ku, Tokyo 112-8610, Japan\\
	${}^{4}$Tokyo Bunkyo SC, the Open Universtiy of Japan, Tokyo 112-0012, Japan \\
	${}^{5}$Center for Neutrino Physics, Department of Physics,\\Virginia Tech,
	Blacksburg VA 24061, USA\\
	${}^{6}$Program for Leading Graduate Schools, \\
	Ochanomizu University, 2-1-1 Ohtsuka, Bunkyo-ku, Tokyo 112-8610, Japan\\
	${}^{7}$Department of Physics, National Tsing Hua University, Hsinchu, Taiwan 300\\
	${}^{8}$Physics Division, National Center for Theoretical Sciences, Hsinchu, Taiwan 300\\
	${}^{*}$bamba@sss.fukushima-u.ac.jp,
	${}^{\dagger}$neil.barrie@ipmu.jp,
	${}^{\ddagger}$sugamoto.akio@ocha.ac.jp,
	${}^{\S}$takeuchi@vt.edu,
	${}^{\P}$kimikoy@phys.nthu.edu.tw}\\
\textit{\small{}}
\par\end{center}{\small \par}

\begin{center}
\textbf{\large{}Abstract}
\par\end{center}{\large \par}

\noindent 
We propose a new non-thermal Leptogenesis mechanism that takes place during the reheating epoch, and utilizes the Ratchet mechanism. 
The interplay between the oscillation of the inflaton during reheating and a scalar lepton leads to a dynamical system that emulates the well-known forced pendulum. 
This is found to produce driven motion in the phase of the scalar lepton which leads to the generation of a non-zero lepton number density that is later redistributed to baryon number via sphaleron processes. 
This model successfully reproduces the observed baryon asymmetry, while simultaneously providing an origin for neutrino masses via the seesaw mechanism.

\end{titlepage}
%%%%%%%%%%%%%%%%%%%%%%%%%%%%%%%%%%%%%%%%%%%%%%%%%%%%%%%%%%%%%%%%%%%%%%%%%%%%%%%%%%%%%%%%%%%%%%%%%%%%
\section{Introduction}

One of the major unsolved problems in modern physics is the origin of the observed baryon asymmetry of the universe.   
The size of the baryon asymmetry is parametrized by the asymmetry parameter $\eta_{B}$ \cite{Ade:2015xua},  
\begin{equation}
\eta_B \,=\, \frac{n_B}{s} \,\simeq\, 8.5 \times 10^{-11}\;,
\label{eta_param1}
\end{equation}
where $n_B$ and $s$ are respectively the baryon number and entropy densities of the universe.

Any $\mathcal{CPT}$ conserving model that wishes to generate this asymmetry must satisfy the so called Sakharov conditions \cite{Sakharov}. 
Although the Standard Model of particle physics does so, it is unable to reproduce a large enough asymmetry, and hence new physics is required. 
It is usually assumed that the baryon asymmetry at the end of the inflationary epoch was negligibly small or zero, due to the rapid dilution of any initial baryon number density that may have existed. 
Due to this, most mechanisms of Baryogenesis are assumed to occur after inflation; during the reheating or subsequent epochs prior to Big Bang Nucleosynthesis.

In what follows we shall outline a new mechanism for Leptogenesis in which lepton number generation is driven by the oscillations of the inflaton field. 
Leptogenesis is a widely studied paradigm that was first suggested in Ref.~\cite{Yanagida:1979as,Fukugita:1986hr}, in which  the baryon asymmetry is proposed to have originated in the leptonic sector. 
Once the lepton asymmetry is generated it is converted to baryon number via $B+L$ violating sphaleron transitions which are in thermal equilibrium prior to the electroweak phase transition \cite{Kuzmin:1985mm,Cohen:1993nk,Trodden:1998ym}. 
See also Ref.~\cite{sugamoto}.

%\cite{Albrecht:1982mp,Kofman:1994rk,Shtanov:1994ce,Kofman:1996mv,Kofman:1997yn,Allahverdi:2010xz}. 

% \cite{Dimopoulos:1978kv,Affleck:1984fy,Cohen:1988kt,Dolgov:1994zq,Dolgov:1996qq,Yoshimura:2004wj,Bamba:2006mh,Minamizaki:2007kj}.

The new Leptogenesis mechanism we propose here acts during the reheating epoch, and is inspired by the ratchet models that describe molecular motors in biological systems \cite{Rachet} and the potential application of it to Baryogenesis \cite{Takeuchi:2010tm}.  In a previous work \cite{Bamba:2016vjs} we considered a toy model consisting of a scalar baryon and inflaton, embedded in the ratchet framework, which aimed to successfully generate the observed baryon asymmetry. Here, we explore more deeply this mechanism from the perspective of Leptogenesis, providing a source for  Baryogenesis and simultaneously providing an origin for the neutrino masses.

%-------------------------------------------------------------------%
%-------------------------------------------------------------------%
%-------------------------------------------------------------------%

%%%%%%%%%%%%%%%%%%%%%%%%%%%%%%%%%%%%%%%%%%%%%%%%%%%%%%%%%%%%%%%%%%%%%%%%%%%%%%%%%%%%%%%%%%%%%%%%%%%%
\section{Description of the Model}

We construct a model consisting of two scalar fields -- a real scalar field $\Phi$ that we identify as the inflaton, and a complex scalar lepton $\phi$. Complex scalars were first utilised for the purposes of Baryogenesis in the Affleck-Dine mechanism \cite{Affleck:1984fy}. In the ensuing analysis we assume that the dynamics during reheating are dictated by these two scalars and  only consider interactions of the inflaton with Standard Model fields via an effective friction term $ \Gamma $ that fixes the reheating temperature. The scalar lepton $ \phi $ also has a friction term associated with its decay to right handed neutrinos. The model is described by the following action:
\begin{eqnarray}
S & = & \int d^4x \,\sqrt{-g}\,
\biggl[\,
g_{\mu\nu}\,\partial^{\mu}\phi^* \partial^{\nu}\phi
\,-\,V(\phi, \phi^*)  
\cr
& & \qquad\qquad\qquad 
+\, \frac{1}{2}\,g_{\mu\nu}\,\partial^\mu\Phi\,\partial^\nu\Phi
\,-\,U(\Phi) \,+\, 
\frac{i}{\Lambda}g_{\mu\nu} \bigl( \phi^*\overleftrightarrow{\partial^\mu}\phi \bigr)
\partial^\nu\Phi 
\,\biggr]
\;,
\label{eq:action}
\end{eqnarray}
where $U(\Phi)$ is the inflationary potential, and $ V(\phi, \phi^*) $ is the scalar lepton potential. The form of the interaction between $\phi$ and $\Phi$ is analogous to that used as in the Baryogenesis mechanism considered in Ref. \cite{Cohen:1987vi,Cohen:1988kt,Arbuzova:2016qfh}. This interaction term is suppressed by the cutoff scale $\Lambda$. In the absence of the $V(\phi,\phi^*)$ term, this action is invariant under a $U(1)$ symmetry which we identify with lepton number, under which $\phi$ has charge 2. The potential $V(\phi,\phi^*)$ is assumed to include a term which breaks this symmetry explicitly.

A key ingredient of our mechanism is the dynamics of the inflaton during reheating. To set up pendulum like dynamics we require that the inflationary potential approaches an $ m^2\Phi^2 $ like potential during reheating. There are various inflationary models that exhibit this behaviour, including the well-known Starobinsky inflationary scenario which is in good agreement with current observational constraints \cite{Ade:2015xua,Starobinsky:1980te}. For illustration purposes, we will discuss our mechanism within the context of the Starobinsky inflationary scenario\footnote{In the Starobinsky context, the introduced derivative coupling term is analogous to that considered in \cite{Davoudiasl:2004gf}.  It has recently been shown that such a coupling may be incompatible with cosmological observations \cite{Arbuzova:2016cem,Arbuzova:2017vdj}.}. In this case, we have the following inflationary potential,
\begin{equation}
U(\Phi)
\;=\; \frac{3\mu^2M_p^2}{4}\Bigl(1-e^{-\sqrt{2/3}\Phi/M_p}\Bigr)^2
\;=\; \frac{1}{2}\mu^2\Phi^2 \,+\, \cdots
\label{Ch3eq:inf_potential1}
\end{equation}
where $\mu = (1.3 \times 10^{-5}) M_p$ is the inflaton mass, and $ M_p = 2.4\times 10^{18}$ GeV is the reduced Planck mass. The reheating period in the Starobinsky model is defined by an $\frac{1}{2}\mu^2\Phi^2 $ potential, leading to the epoch being characterised by a time averaged Hubble rate that is analogous to the Hubble rate of a matter dominated epoch \cite{Kofman:1996mv}.  The reheating epoch in this scenario is characterised by the following initial parameters, from numerical calculations: $ \Phi_i = \Phi(t_i) = 0.62\,M_p $,
with a corresponding Hubble parameter of
$ H_i = H(t_i) =   6.2\times 10^{12}~\mathrm{GeV} $ \cite{Ellis:2015pla}.

We consider the potential associated with the scalar lepton to 
include an explicit lepton breaking term of the following form\footnote{%
Through introducing an additional scalar lepton $\varphi $, we may naturally realize a potential of this form via  spontaneous symmetry breaking $\langle\varphi\rangle\neq 0$. In order for this mechanism to work, however, the Nambu-Goldstone boson associated with this spontaneous breaking must be eliminated.
}:
\begin{eqnarray}
V(\phi, \phi^*) \;=\; V_0\left(|\phi|^2\right)
-\lambda\,\phi\phi^*(\phi-\phi^*)^2 \;.
\label{Ch3eq:b_potential0a} 
\end{eqnarray}
Hence, the action during reheating is,
\begin{eqnarray}
S 
& = & \int d^4x \,\sqrt{-g}\,
\biggl[\,
g_{\mu\nu}\,\partial^{\mu}\phi^*\partial^{\nu}\phi
\,+\,\lambda\phi\phi^*(\phi-\phi^*)^2
\cr 
& & \qquad\qquad\qquad 
+\,\frac{1}{2}\,g_{\mu\nu}\,\partial^\mu\Phi\,\partial^\nu\Phi
\,-\,\frac{1}{2}\mu^2\Phi^2 
\,+\,\frac{i}{\Lambda}g_{\mu\nu} 
\bigl( \phi^*\overleftrightarrow{\partial^\mu}\phi \bigr)\partial^\nu\Phi 
\,\biggr]
\;,
\label{Ch3eq:actiona}
\end{eqnarray}
where we have neglected terms associated with  $ V(|\phi|^2) $ which will not be important for our analysis.
It is clear that if $\lambda=0$, the action will be invariant under the
global $U(1)_L$ symmetry defined by the transformation 
$(\phi, \phi^*) \to (e^{-2i\alpha}\phi, e^{2i\alpha}\phi^*)$, 
where $\alpha$ is a constant. 
This transformation has the corresponding lepton number current, 
\begin{equation}\label{current}
j_{L}^{\mu}
\;=\; -2i\left( \phi\, \partial^{\mu}\phi^{*} -\phi^{*}\partial^{\mu}\phi \right) 
\,+\,\frac{4|\phi|^2}{\Lambda}\,\partial^\mu\Phi \;.
\end{equation}
%	The coupling between the inflaton and $ \phi $ can then be seen as the  $\mathcal{L}_{int}=-\frac{1}{v} j_L^{\mu} \partial_{\mu} \Phi$. This interaction term violates $\mathcal{C}$ and $\mathcal{CP}$, which is a necessary ingredient for successful Leptogenesis.
%
We now wish to consider the following polar coordinate parametrization of the $\phi$ field,
$ \phi = \frac{1}{\sqrt{2}}\phi_r e^{i\theta}  $. Under the global lepton number  transformation, the phase $\theta$ transforms as
$\theta \rightarrow \theta-2\alpha$, 
while $\phi_r$ is invariant.
In this parametrization the conserved lepton number density, which corresponds to the time component of 
Eq.~\eqref{current}, is given by,
\begin{equation}
\label{Ch3eq:baryon_density_calc}
n^{\prime}_L \;=\; j^0 \;=\; -2\phi_r^2\biggl(\dot{\theta} -\frac{\dot{\Phi}}{\Lambda}\biggr)\;.
\end{equation}
The non-conserved physical net lepton number density is that from the free-field Lagrangian,
\begin{equation}
\label{Ch3eq:baryon_density_calc2}
n_L \;=\; - 2\phi_r^2 \dot{\theta}\;.
\end{equation}
This implies that within the framework of our mechanism we must produce a non-zero $\dot{\theta}$, a period of driven motion, to have a net lepton asymmetry generated. In the rest of our analysis, we assume that the terms that only depend on $\phi_r$ in $V$ are such that they keep $\phi_r$ approximately fixed to a constant non-zero value, and that only the dynamics of the phase $\theta$ need be considered.

%		The charge conjugation symmetry $\mathcal{C}$ is given by $\mathcal{C}: \phi \to \phi^*$, or $\theta \to -\theta$, so $\mathcal{C}$ is conserved in this potential.  $L$ invariance is related to the translational invariance of the phase $ \theta $, which is clearly violated by this potential, assuming $\lambda\neq 0$.

Seeing as we wish to consider the cosmological setting of reheating, 
we take the flat FRW metric with scale factor $ a(t) $.
Given this isotropic and homogeneous background, we extend this assumption to the properties of the scalar lepton and inflaton, for which spatial variation will be ignored in our analysis. Therefore, in the new parametrization of the scalar lepton and in a flat FRW background, the action takes the form,
\begin{eqnarray}
S = 
\int dt  \; a(t)^3
\left[\,
\frac{\phi_r^2}{2}\,\dot{\theta}^2
\,-\, \lambda\,\phi_r^4 \sin^2\theta
\,+\, \frac{1}{2}\,\dot{\Phi}^2
\,-\, \frac{1}{2}\,\mu^2\Phi^2 
\,-\, \frac{\phi_r^2}{\Lambda}\,\dot{\theta}\,\dot{\Phi} 
\,\right]\;.
\label{Ch3eq:action2a}
\end{eqnarray}
This action illustrates how the Sakharov conditions are satisfied in our model. Firstly,
$L$ violation is achieved by the potential $V_{\textrm{int}}=\lambda\phi_r^4 \sin^2\theta$,
which breaks the translational invariance in $\theta$.
Secondly, the derivative coupling between $ \theta $ and $ \Phi $ provides $\mathcal{C}$ and $\mathcal{CP}$ violation. Lastly, the required push out-of-equilibrium will be provided by the reheating epoch, induced by the coherent oscillation of the inflaton field. The lepton number asymmetry generated during reheating shall then be redistributed into a net baryon number by the action of $B-L$ conserving sphaleron processes \cite{Kuzmin:1985mm,Trodden:1998ym,sugamoto}.

The generated lepton asymmetry will be produced in the form of right handed neutrinos, via the preferential decay of $ \phi $ during the period of driven motion. To achieve this, we introduce the lepton-number preserving dimension four interactions,
\begin{equation}
\Delta\mathcal{L}_{\mathrm{int}} \;=\; 
\left(\,
g_{L}\phi^* \overline{\nu_R^c}\nu_R^{\phantom{c}} 
\,+\, y_H H\overline{L} \nu_R 
\,\right)
\,+\, \mathrm{h.c.}
\label{nu_int}
\end{equation}
which describes the coupling of the $ \phi $ field to the right handed neutrinos. 
We add also the Standard Model Higgs field, $H$, and left handed lepton doublet, $L$, 
through their Yukawa coupling to the right handed neutrino. 
This interaction shall play a role in the generation of the active neutrino masses in this model.
The interaction term containing  $ \phi $ is responsible for the generation of the right handed neutrino mass. This shall take part in the neutrino mass generating model known as the
seesaw mechanism \cite{Yanagida:1979as,
%Yanagida:1980xy,%
Ramond:1979py}, which gives an explanation for the small masses of the active neutrinos.

%During reheating the derivative coupling between the scalar lepton and inflaton can lead to a kind of resonance effect when the respective potentials are of a similar order. As we shall see, when this is the case we can observe directed motion in the scalar lepton phase, and hence a non-zero lepton number density. 		

Now, we can carry out the calculations required to determine the lepton number density generated in this framework. 

%In our analysis we will take the initial phase of the scalar lepton to be zero, 
%placing the scalar lepton initially in the minimum of its potential. 
%To ensure there is no initial bias between matter and antimatter we assume that the initial phase velocity $ \dot{\theta} $ is zero.

%-------------------------------------------------------------------%
%-------------------------------------------------------------------%
%-------------------------------------------------------------------%

%%%%%%%%%%%%%%%%%%%%%%%%%%%%%%%%%%%%%%%%%%%%%%%%%%%%%%%%%%%%%%%%%%%%%%%%%%%%%%%%%%%%%%%%%%%%%%%%%%%%
\section{Analysis of the Period of Driven Motion}

We shall now find an analytical solution for the equation of motion of the scalar lepton phase $ \theta $  to determine the region of parameter space where we obtain driven motion, and hence can produce a non-zero lepton number density. Firstly, we find the equations of motion for $ \Phi $ and $ \theta $ using the action presented in Eq.~\eqref{Ch3eq:action2a},
\begin{eqnarray}
\label{Ch3eq:diff_eq}
\left( \ddot{\Phi} + 3H\dot{\Phi} \right) + 
\left( \Gamma\dot{\Phi} + \mu^2\Phi \right) -\, 
\frac{\phi_r^2}{\Lambda}
\left(\ddot\theta + 3H\dot\theta \right) 
& = & 0\;, 
%\vphantom{\bigg|}
\\
\left( \ddot\theta + 3H\dot\theta \right) +\,
%+ \lambda\phi^2_r\sin(2\theta)
\lambda\,\phi_{r}^2\sin(2\theta) \,-\,
\frac{1}{\Lambda}
\left( \ddot\Phi + 3H\dot\Phi \right)  & = & 0\;,
%\vphantom{\bigg|}
\end{eqnarray}
%
%These coupled differential equations 
which can be simplified, assuming $\phi_r^2/\Lambda^2 \ll 1 $, to
\begin{eqnarray}
\left( \ddot{\Phi} + \frac{2}{t}\dot{\Phi} \right) +
\left( \Gamma\dot{\Phi} + \mu^2\Phi \right) +\, 
\frac{\lambda\phi^4_r}{\Lambda}\sin (2\theta)
& = & 0\;, 
%\vphantom{\bigg|} 
\label{Ch3PhiEq}
\\
\left( \ddot\theta + \frac{2}{t}\dot\theta \right) +\,  
\frac{1}{\Lambda}
\left( \Gamma\dot{\Phi} + \mu^2\Phi \right) +\, 
\lambda\,\phi_{r}^2\sin(2\theta)  
& = & 0\;, 
%\vphantom{\bigg|}
\label{Ch3PhithetaEq}
\end{eqnarray}
where $\Gamma\dot{\Phi}$ is the inflaton friction term, added in by hand, which encapsulates the decay of the inflaton and fixes the reheating temperature.

%After some rearrangement, the above equations read,
%\begin{eqnarray}
%\left(1- \frac{\phi_r^2}{v^2}\right)
%(\ddot{\Phi} + 3H\dot{\Phi}) + \left(\Gamma\dot{\Phi} + \frac{dU(\Phi)}{d\Phi} \right)
%+ \lambda\phi^2_r v\sin (2\theta)
%& = & 0~, \vphantom{\bigg|}
%\\
%\left(1- \frac{\phi_r^2}{v^2}\right)
%\bigl(\ddot\theta + 3H\dot\theta\bigr)
%+  \frac{1}{v}
%\left(\Gamma\dot{\Phi} 
%+  \frac{dU(\Phi)}{d\Phi} 
%\right)
%+ \lambda v^2\sin(2\theta)  
%& = & 0~. \vphantom{\bigg|}
%\end{eqnarray}
%The $ \left(1- \frac{\phi_r^2}{v^2}\right) $ factor can be neglected, as we shall require that $ \frac{\phi_r^2}{v^2} \ll 1$. 

%It is also assumed that during reheating the Starobinsky potential can be expressed approximately as,	
%\begin{equation}
%U(\Phi) 
%\approx  \frac{1}{2} \mu^2\Phi^2 
%\quad\rightarrow\quad
% \frac{d U(\Phi)}{d\Phi} \approx  \mu^2\Phi~,
%\end{equation}
%when taking $ \Phi\ll M_p $. In such a potential, the oscillation of the inflaton gives rise to an approximate matter dominated epoch, during which the Hubble parameter $H$ is given by,
%\begin{equation}
%\label{Ch3eq:time_hubble}
%H = \frac{\dot{a}}{a} \approx \frac{2}{3t} ~.
%\end{equation}

% Note that early on in the reheating epoch the full potential should be used. This approximation is valid towards the end of the epoch, when the inflaton is oscillating near the potential minimum.
%The equations of motion are simplified as follows,

%\subsubsection*{Behaviour of the Inflaton}
We wish for the inflaton's motion to be unaffected by the dynamics of $\theta$. This is to ensure that the properties of the reheating epoch and the coherent oscillation of the inflaton are retained.
To do so, we assume that the $\sin(2\theta)$ term in
Eq.~\eqref{Ch3PhiEq} can be neglected. The equation of motion for the inflaton becomes,
\begin{equation}
\ddot{\Phi} \,+\, \left( \frac{2}{t} +  \Gamma \right)\!\dot{\Phi} \,+\,  {\mu}^2\Phi \;=\; 0\;.
\end{equation}
This equation can be easily solved in the case when $ \Gamma\ll  \mu$, which is a valid assumption in our scenario.
The approximate solution to this equation is,
\begin{equation}
\Phi(t)
\;\simeq\; \Phi_i \left( \frac{t_i}{t}\right)
e^{-\Gamma(t-t_i)/2}\cos\bigl[ \mu(t-t_i)\bigr] \;,
\label{Ch3PhiSolutionT}
\end{equation}
where $t_i$ is the time at which the reheating epoch begins, and $\Phi_i=\Phi(t_i)$.
This solution indicates that the motion of $\Phi(t)$ is oscillatory, with an angular frequency $ \mu$, and an  amplitude predominantly attenuated  by Hubble damping early in reheating.

Now that we have this solution it is possible to find a simple relation describing the assumption that the $\sin(2\theta)$ term can be neglected in the equation of motion
of $\Phi$. This requires that the following relation be satisfied,
\begin{equation}
\frac{\lambda\phi^4_r}{\Lambda} \;\ll\; \mbox{amplitude of $ \mu^2\Phi(t)$}\;.
\label{Ch3AmplitudeCondition1}
\end{equation}
This should be true throughout the reheating epoch, and will be discussed further below.

%%%%%%%%%%%%%%%%%%%%%%%%%%%%%%%%%%%%%%%%%%%%%%%%%%%%%%%%%%%%%%%%%%%%%%%%%%%%%%%%%%%%%%%%%%%%%%%%%%%%
%\subsubsection*{Behaviour of the Scalar Lepton}
Now that we have determined the dynamics of the inflaton during reheating, we can find an analytical solution for the phase of the scalar lepton. As found above,
the equation of motion for $\theta$ is,
\begin{eqnarray}
\left(\ddot{\theta} + \frac{2}{t}\dot{\theta}\right)
\,+\, p\sin(2\theta)
\,+\, q(t)\cos\bigl[ \mu(t-t_i)\bigr]
\;=\; 0\;,
\label{Ch3thetaeom}
\end{eqnarray}
where the inflaton decay term has been dropped, as the amplitude of $ \Gamma\dot{\Phi}(t)$ is
suppressed compared to the amplitude of $ \mu^2\Phi(t)$ due to $\Gamma \ll \mu$, 
and we have defined
\begin{equation}
p\,=\,\lambda\phi_{r}^2\;,\qquad  
q(t)\,=\,\dfrac{\mu^2\Phi_i}{\Lambda}\frac{H(t)}{H_i}\;.
\end{equation}
Eq.~\eqref{Ch3thetaeom} is not simple to solve, so we shall first consider a few possible scenarios to determine some of its properties.	
First, consider the case when $p\ll q(t)$, we now have,
\begin{equation}
\left(\ddot{\theta}+ \frac{2}{t}\dot{\theta}\right)
\;=\;  \frac{1}{t^2} \frac{d}{dt}\left(t^2\dot{\theta}\right)
\;=\; -q(t)\cos\bigl[ \mu(t-t_i)\bigr]
\;,
\end{equation}
which can be integrated to yield,
\begin{eqnarray}
\dot{\theta}(t)
\;=\; 
-\left( \frac{\Phi_i}{ \Lambda}\right)
\frac{t_i}{t^2}
\Bigl(
\cos\bigl[ \mu(t-t_i)\bigr]
+ \mu t\sin\bigl[ \mu(t-t_i)\bigr]
\Bigr)
\;=\; \frac{\dot{\Phi}}{\Lambda}
\;.
\end{eqnarray}
Immediately we can see that there is no  lepton number violation in this case, by comparing this to the leptonic current presented in Eq.~\eqref{Ch3eq:baryon_density_calc}. 
We find that there is no dependence on $ \theta $, specifically evidence of the  $\lambda \phi_{r}^2 \sin(2\theta)$ term is absent. Thus, we can see that, in this limit
the motion of $\dot{\theta}$ is driven solely by the oscillation of the inflaton and
simply oscillates around zero, not maintaining any finite value.
This is to be expected since this limit is equivalent to removing the $L$ violating term associated with the scalar lepton potential.

Now consider the limit $p\gg q(t)$, for which the equation of motion becomes,
\begin{equation}
\ddot{\theta} \,+\, \frac{2}{t}\dot{\theta} \,+\, p\sin(2\theta)
\;=\; 0\;.
\end{equation}
In this case, if we start from a state with finite energy,
the friction term will damp the motion of the phase $\theta$ until 
it settles into one of its potential minima, and again there will be
no non-zero $\dot{\theta}$ which persists and can lead to a non-zero lepton number density.
Of course, this is to be expected since in this limit the $\mathcal{C}$ and $\mathcal{CP}$
breaking term has been removed.

Therefore, we can conclude that for successful asymmetry generation, we require $p\simeq q(t)$ so that
both the $L$ breaking and the $\mathcal{C}$ and $\mathcal{CP}$ terms can contribute
to the time evolution of $\theta$. Thus, during reheating we must achieve $p\simeq q(t_d)$ at some time $ t_{d} $, which we shall name the Sweet Spot Condition (SSC):
\begin{equation}
\underbrace{\vphantom{\bigg|}\;\lambda \phi_{r}^2\;}_{\displaystyle p} 
\simeq\; \underbrace{\frac{\mu^2\Phi_i}{\Lambda} \left( \frac{H_d}{H_i} \right)}_{\displaystyle q(t_d)}
\;=\;      \frac{\mu^2\Phi(t)}{\Lambda} \left( \frac{H_d}{H(t)} \right) 
\;,
\label{Ch3FinalTimeCondition2}
\end{equation}
where $\Phi(t)/H(t)$ is constant during the matter dominated epoch. In what follows, we will associate $ \Lambda $  with the GUT scale, and assume $\Lambda=10^{16}$~GeV.

% Before continuing the analysis we should comment on the scales $ \Lambda $ and $ v $, given the  dependence of the SSC upon them. The scale $v$ is related to both the mass of the right handed neutrino, and the scale of lepton number violation, while the scale $\Lambda$ is related to the breaking scale of $ C $ and $ CP $.  On considering the derivative coupling term associated with the scale $ \Lambda $, presented in Eq. (\ref{eq:action}), it is observed to be related to the leptonic current. With this in mind we assume that the scales $ \Lambda $ and $ v $ are related and satisfy the following relation,
%\begin{eqnarray}
%\Lambda=\sqrt{2} \langle \varphi \rangle =v~,
%\label{vev_rel}
%\end{eqnarray}
%in what follows, we will associate $ \Lambda $ and $ v $ with the GUT scale, and assume $ \Lambda = v =10^{16} $ GeV.

%%%%%%%%%%%%%%%%%%%%%%%%%%%%%%%%%%%%%%%%%%%%%%%%%%%%%%%%%%%%%%%%%%%%%%%%%%%%%%%%%%%%%%%%%%%%%%%%%%%%
\subsection{Phase Locked States and the Forced Pendulum}

A rigorous solution for $ \theta $ can be found by drawing an analogy between our mechanism and a forced pendulum.
In Eq.~\eqref{Ch3thetaeom}, the term proportional to $\sin(2\theta)$ can be viewed as the gravitational force on the pendulum, when it is at an angle $2\theta$ from the vertical, 
$q(t)$ the external pushing force, and a friction term 
$f(t)= 3H+ \Gamma_{\phi} =(2/t) + \Gamma_{\phi} $, 
where we have included the decay width to right handed neutrinos denoted $\Gamma_{\phi}$.  There is an added complexity in our case, in that the strength of the external force $q(t)$ and the friction $f(t)$ on the pendulum both depend on $t$. The time evolution of $q(t)$ is expected to be slow relative to the frequency of the driving force $ \mu $, that is $ H\ll\mu $, so to analyse the dynamics of $\theta$ within that time frame, it is sufficient to replace it with the
constant $q(t_d)$, similarly with the Hubble friction term. It shall be assumed that 
$\Gamma_{\phi}\gtrsim H_d = H(t_d)$, such that during the period of driven motion the production of right handed neutrinos is the dominant source of friction.

In order to produce driven motion, the timing and intensity of the external push must match the motion of the pendulum, which is the idea embodied by the SSC, $p\approx q(t_d)$. If this is satisfied, the rotational motion of the pendulum around the fixed point arises with an almost constant angular velocity $\dot{\theta}$.

The relevant solutions to the the equation of motion in our scenario are 
those that increase or decrease monotonously in time with only small amplitude modulations.  
Such solutions exist and are known as \textit{phase-locked states}, which are found in the study of  the 
chaotic behaviour of the forced pendulum.  
The conditions for phase-locked states to exist were considered in the study of chaotic behaviour of electric current passing through a Josephson junction \cite{Pedersen:1980}.
We shall  follow the notation adopted in these studies \cite{D'Humieres:1982}, and change the variables as follows,
\begin{equation}
\Theta \,\equiv\, 2\theta\,,\quad
\tau   \,\equiv\, \sqrt{2p}\left[(t-t_i)- \frac{\pi}{\mu}\right],\quad
\omega \,\equiv\, \frac{\mu}{\sqrt{2p}}\,,\quad
Q      \,\equiv\, \frac{\sqrt{2p}}{3 H_d+\Gamma_{\phi}}\,,\quad
\gamma \,\equiv\, \frac{q(t_d)}{p}\,.
\end{equation}
Thus, the equation of motion becomes,
\begin{equation}
\ddot{\Theta} \,+\, \frac{1}{Q}\dot{\Theta} \,+\, \sin\Theta \;=\; \gamma\cos(\omega \tau)\;.
\label{Ch3EOM}
\end{equation}
Our equation coincides exactly with that of the forced pendulum or Josephson junctions. The generic phase-locked state solution to the above equation has the following form, when $ \gamma\approx 1 $,
\begin{eqnarray}
\Theta(\tau) \;=\; 
\Theta_0 \,+\, n \omega \tau \,-\, \sum_{m=1}^{\infty} \alpha_m \sin (m \omega \tau -\delta_m)\;,
\label{THETAgeneric}
\end{eqnarray}
where $n$ and $m$ are integers.
In the numerical calculations we performed only the phase-locked states with $m=1$ appear.  In such solutions the period of the amplitude modulation is equal to that of inflaton's oscillation.  Hence the solution to our equation of motion is of the form,
\begin{equation}
\Theta \;=\; \Theta_n \,+\, n (\omega \tau-\delta) \,-\, \alpha \sin (\omega \tau -\delta)\;.
\label{THETAspecific}
\end{equation}
For these solutions, we can calculate the lepton number density $n_L$ as  
the time average of $\dot{\Theta}$. From this we arrive at the following,
\begin{equation}
n_L 
\;=\; -2\phi_r^2\langle\dot{\theta}\rangle
\;=\; -2\sqrt{\frac{p}{2}}\,\phi_r^2\langle\dot{\Theta}\rangle 
\;=\; -2\sqrt{\frac{p}{2}}\,\phi_r^2n \omega 
\;=\; -\left(\mu \phi_r^2\right)n
\;.
\label{Ch3eq:barnodenEstimate}
\end{equation} 
Interestingly, this result depends on the integer $n$, 
where $n/2$ is the number of rotations of the phase $\theta$ 
per oscillation of the inflaton.  
The value of $n$ is not given by the solution and hence we must determine it using numerical simulations. In the following section we attempt to obtain an approximation for the value of $ n $.

%%%%%%%%%%%%%%%%%%%%%%%%%%%%%%%%%%%%%%%%%%%%%%%%%%%%%%%%%%%%%%%%%%%%%%%%%%%%%%%%%%%%%%%%%%%%%%%%%%%%
\subsection{Approximate Analytical Solution and $n$}

Starting with the equation of motion for $ \theta $ in the form given in Eq.~\eqref{Ch3thetaeom} where the friction term has been dropped for simplicity, and it is assumed that we are in the regime consistent with the SSC,
\begin{eqnarray}
\ddot\theta 
\;+\; \lambda \phi_{r}^2
\bigl[\,\sin(2\theta) + \cos(\mu t)\,\bigr]
\,=\; 0\;.
\end{eqnarray}
Make the following reparametrization, $ \tau=\mu t $, under which derivatives are denoted by primes, 
$ \xi= \mu^2 \theta/\lambda \phi_{r}^2 $ and $ n = 2\lambda \phi_{r}^2/\mu^2$. This gives,
\begin{eqnarray}
\xi^{\prime\prime}
\,+\, \sin\left(n\xi\right)
\,+\, \cos(\tau)
\;=\;0\;.  
\label{the original EOM} 
\end{eqnarray}
%
%Comparing the result of the approximate solution with that of the phase-locked state solution gives the following value $ n=\frac{2\lambda v^2}{\mu^2} $. Interestingly this value  naturally appears in the equations above, from the modulation of the frequency of the $ \sin $ function relative to that of the driving force of the inflaton denoted by the $ \cos $ function. To simplify we shall use this notation,
%
Now we shall assume the following ansatz, that directed motion is present 
$\xi=\tau+\,\cdots$, and match this to the phase locked state solution for $n$ and 
$\langle\dot\theta\rangle$,
where the ellipses represent initial phases and oscillatory terms which go to zero upon averaging $\xi$. That is, reinterpreting this as $\langle\dot\theta\rangle$ we obtain,
\begin{eqnarray}
\langle\dot\theta\rangle \;=\; \dfrac{n\mu}{2}\;.
\end{eqnarray}
We shall also assume that these oscillating terms are not dominant over the directed motion term, and see if this is consistent with the simplified equation of motion which is exactly solvable,
\begin{eqnarray}
\xi^{\prime\prime}
\,+\, \sin(n\tau)
\,+\, \cos(\tau)
\;=\; 0\;.
\label{eqm}
\end{eqnarray}
This equation of motion is easily solved and has the general solution,
\begin{eqnarray}
\xi \;=\; a 
\,+\, b\tau
\,+\, \frac{\sin(n\tau)}{n^2}
\,+\,\cos(\tau)
\;.
\end{eqnarray}
In order for this solution to be consistent with our ansatz, and the original equation of motion in Eq.~\eqref{the original EOM},  $b$ must  be $1$, upon taking the derivative and time averaging.
Hence, it is found that,
\begin{eqnarray}
\langle\xi^{\prime}\rangle \;=\; 1\;,
\end{eqnarray}
which implies that the parametrization of $n$ considered here is consistent with that given in the phase-locked solution.
This can be easily verified by solving Eq.~\eqref{Ch3thetaeom} numerically, which confirms the consistency of the approximation,
\begin{eqnarray}
n \;=\; \frac{2\lambda \phi_{r}^2}{\mu^2}\;.
\label{approx_n}
\end{eqnarray}
Using this value of $ n $ we can now proceed with calculating the asymmetry density generated by the driven motion in $ \theta $. That is,
\begin{equation}
|n_L| \;\approx\; \frac{2\lambda \phi_{r}^4}{\mu}\;.
\label{Ch3eq:barnoden}
\end{equation}

%%%%%%%%%%%%%%%%%%%%%%%%%%%%%%%%%%%%%%%%%%%%%%%%%%%%%%%%%%%%%%%%%%%%%%%%%%%%%%%%%%%%%%%%%%%%%%%%%%%%
\subsection{Dynamics after Driven Motion}

Once the SSC is violated there is no net production of  $L$, as simultaneous violation of $\mathcal{C}$, $\mathcal{CP}$ and $L$ will not be realized. 
The phase $\theta$ becomes constrained around a singular minimum around which it oscillates due to the motion of the inflaton, and damped by the friction term defined by $\Gamma_{\phi}$. To describe this period we take small $\theta$ and $H \ll H_{d}$, to allow the approximation of the amplitude of these oscillations,
\begin{eqnarray}
\theta^{\prime \prime}
\,+\, \frac{\Gamma_{\phi}}{\mu}\theta^{\prime}
\,+\, n \theta
\,+\, \frac{n}{2}\frac{H(\tau)}{H_{d}}\cos(\tau) 
\;=\; 0\;,
\label{thetaeom}
\end{eqnarray}
where $n = 2\lambda \phi_{r}^2/\mu^2$, and $ \tau=\mu t $, under which primes denote derivatives. An approximate solution to this equation can be found when taking $ H\ll H_{d} $ and $ \Gamma_{\phi}\ll \mu $,
\begin{equation}
\theta \;\approx\; \left(\frac{n}{n-1} \right)\frac{H}{2 H_{d}} \cos(\tau) \;,
\end{equation}
where this gives an upper limit on the oscillation amplitude,
\begin{equation}
\theta_\mathrm{max} \;\approx\; \left(\frac{n}{n-1} \right) \frac{H}{2H_d}\;.
\label{max_amp}
\end{equation}	
This result can now be used to check the consistency of our previous assumptions, and provide additional constraints on the model parameters. It should also be noted that the solution for $ \dot{\theta} $ in this period oscillates around zero with no driven motion, and when time-averaged is zero.

Prior to the period of driven motion, it is easy to see that the dynamics of the inflaton are unaffected by the dynamics of $ \theta $. From the above result, in Eq.~\eqref{max_amp}, we can also consider the requirements for this to be the case after driven motion has occurred. Firstly, the maximum scalar lepton energy density   should not exceed that associated with the inflaton field,
\begin{equation}
3M_p^2H^2 \;\gg\; \lambda \phi_{r}^4 \sin^2\theta \;,
\end{equation}
utilizing the SSC and Eq.~\eqref{max_amp}, this gives,
\begin{equation}  
\lambda \;\gg\; 8\times 10^{-6}\;,
\label{con_1}
\end{equation}
where we have assumed that for most cases we can use $ (n-1)/n \approx 1 $.

The second necessary condition is that the equation of motion of $ \Phi $, given in Eq.~\eqref{Ch3PhiEq}, is unaffected by $ \theta $ during reheating:
\begin{equation}
\mu^2 \Phi(t) \;\gg\; \frac{\lambda\phi_r^4}{\Lambda}\sin(2\theta) \;.
\end{equation}
Utilizing the SSC we obtain,
\begin{equation}
\frac{\phi_r^2}{\Lambda^2} \;\ll\; 1 \;,
\label{limit}
\end{equation}
which is consistent with the assumption we made to derive the equations of motion in 
Eq.~\eqref{Ch3PhiEq} and Eq.~\eqref{Ch3PhithetaEq}. 
In this analysis we assume that the scale $ \Lambda $ corresponds to the GUT scale. Thus, it is necessary that,
\begin{equation}
10^{15}\,\textrm{GeV} \;\gtrsim\; \phi_{r}\;.
\label{con_2}
\end{equation}

We can also apply direct constraints on the Hubble rate at which the driven motion occurs $ H_d $. From the SSC, Eq.~\eqref{Ch3FinalTimeCondition2}, we find that $ H_d $ can be expressed as, 
\begin{equation}
H_d \;=\; 2n\times 10^{10}\,\textrm{GeV}\;.
\end{equation}
Since the driven motion is taking place during reheating, it is necessary that $ H_i > H_d $, so using this form we can apply constraints on the allowed values of $ n $:
\begin{equation}
310 \;>\; n \;>\; 1\;.
\end{equation}
The requirement for driven motion $ n>1 $ allows a lower limit to be placed on $ \lambda $ when considering the condition in Eq.~\eqref{limit} and the approximate value of $ n $,
\begin{equation}
1 \;>\; \lambda \;>\; 5\times 10^{-4}\;,
\label{con_3}
\end{equation}
which means that the condition derived in Eq.~\eqref{con_1} is immediately satisfied in the case of driven motion. 
The upper bound is required for perturbativity, and implies $ \phi_r \geq \mu $ from $ n>1 $.

%From Eq. (\ref{approx_n}), this can be reinterpreted as bounds on $ \lambda $,
%\begin{equation}
%1.5\times 10^{-3} >\lambda>10^{-5}~,
%\end{equation}
%from this it is easy to see that the constraints given in Eq. (\ref{con_1}) and (\ref{con_2}) are consistent and readily satisfied.
%
%\begin{equation}
%\Gamma_{\phi}=\frac{g^2 m_{\phi}}{8\pi}\left(1-\frac{4m_f^2}{m_{\phi}^2}\right)^{3/2}
%\end{equation}

%
%
%Decay width of right handed neutrino to higgs and charged lepton,
%$ \Gamma_{\nu_R}=\frac{y^2 m_{\nu_R}}{8\pi} $
%for $ m_{\nu}=\frac{m_{LR}^2}{m_{\nu_R}}=\frac{y^2 v^2}{m_{\nu_R}} $,
%$ \Gamma_{\nu_R}=\frac{ m_{\nu} m_{\nu_R}^2}{8\pi v^2} $
%where $  y\simeq 0.2 \sqrt{\frac{m_{\nu_R}}{\mu}} $.
%$ \Gamma_{\nu_R}\simeq 6\times 10^{10} c^2 $ GeV, where $ m_{\nu_R}=c \mu $.
%
%

%-------------------------------------------------------------------%
%-------------------------------------------------------------------%
%-------------------------------------------------------------------%

%%%%%%%%%%%%%%%%%%%%%%%%%%%%%%%%%%%%%%%%%%%%%%%%%%%%%%%%%%%%%%%%%%%%%%%%%%%%%%%%%%%%%%%%%%%%%%%%%%%%
\section{Right Handed Neutrino Properties}

Before considering the size of the resultant baryon asymmetry, we shall look more closely at the neutrino sector in this model.  The mass of the right handed neutrino is dictated by the Yukawa interactions between $ \nu_{R} $ and the scalar lepton, $ \phi $, introduced in Eq.~\eqref{nu_int}. It is necessary that at least the lightest right handed neutrino must have a mass much less than that of $ \phi $, so that the decays of $ \phi $ during the driven motion are not kinematically forbidden. From Eq.~\eqref{nu_int}, the right handed neutrino will have the following Majorana mass:
\begin{equation}
m_{\nu_{R}} \;=\; \frac{g_L \phi_r}{\sqrt{2}}\;,
\label{nu_mass}
\end{equation} 
where we assume $ \phi $ gives the dominant contribution to the right handed neutrino mass. Here we will consider the dynamics of a single right handed neutrino, but this can be easily extended to three generations.

In this Leptogenesis scenario, the preferential decay of the $ \phi $ to right handed neutrinos during the period of driven motion is the production mechanism for the lepton asymmetry. In order for this decay to be relevant during this period  it is necessary that $ \Gamma_{\phi}\gtrsim H_d $. The decay width of $ \phi $ to the right handed neutrinos, via this interaction can also be used to apply constraints on the couplings. Given that $ m_{\phi}\gg  2 m_{\nu_R} $, the decay width is of the form,
\begin{equation}
\Gamma_{\phi} \;=\; \frac{g_L^2 m_{\phi}}{8\pi}\;.
\end{equation}
In our analysis we require that $ \Gamma_{\phi}>H_d $, thus we can place a lower limit on the coupling $ g_L $, assuming the mass of $ \phi $ is less than the GUT scale. That is,
\begin{equation}
g_L \;>\; 10^{-2} \;,
\label{con_g}
\end{equation}
Now we can find the allowed range of masses for the right handed neutrino,
\begin{equation}
10^{14}\,\textrm{GeV} \;>\; m_{\nu_{R}} \;>\; 10^{11}\,\textrm{GeV}\;,
\end{equation}
where the upper bound is derived from the requirement that $ m_{\phi}\gg  2 m_{\nu_R} $, and the lower bound is found from combining the lower limit on $ \phi_r $ from $ n>1 $ and $ \lambda<1 $, with Eq.~\eqref{nu_mass} and Eq.~\eqref{con_g}.
%From this decay width and Eq. (\ref{nu_mass}) we can get the following relation between the model parameters,
%\begin{equation}
%\frac{g_L}{\sqrt{2}}\ll\frac{m_{\phi}}{\phi_{r}}~.
%\end{equation}

The right handed neutrinos cannot play a role in the sphaleron transitions, so prior to the redistribution of the lepton asymmetry into a baryon number asymmetry, the right handed neutrinos must decay into the Standard Model leptons. We will assume that this occurs predominately via the Higgs and Standard Model lepton channel described in Eq.~\eqref{nu_int}. The corresponding decay width is given by,
\begin{equation}
\Gamma_{\nu_R} \;=\; \frac{y_H^2 m_{\nu_R}}{8\pi}\;,
\label{nu_decay}
\end{equation}
where $ y_H $ is the Yukawa coupling between the Higgs, right handed neutrino, and left handed lepton doublet. This Yukawa coupling in combination with that between the right handed neutrinos and $ \phi $ constitute the ingredients for the seesaw mechanism, generating a mass for the active neutrinos. In this simplified scenario, the left and right handed neutrino masses are related by the following mass matrix, 
\begin{equation}
\begin{bmatrix}
0     & m_{D}  \\
m_{D} & m_{\nu_R} 
\end{bmatrix}
\end{equation}
obtained from the Dirac mass term $m_{D}$ and the right handed neutrino mass term $m_{\nu_R}$. 
Upon diagonalizing this matrix we find the masses for the active neutrinos given by $m_{\nu_L} \simeq m_{D}^2/m_{\nu_R}$, 
and that associated with the right handed neutrino $ m_{\nu_R} $.
The mass of the active neutrinos is given by,
\begin{equation}
m_{\nu_L} \;=\; \frac{y_H^2 v_h^2}{2 m_{\nu_R}} \;,
\end{equation}
where  $ v_h \simeq 246$ GeV is the Higgs vacuum expectation value. If we combine this with Eq.~\eqref{nu_decay},
\begin{equation}
\Gamma_{\nu_R} \;=\; \frac{m_{\nu_L} m_{\nu_R}^2}{4\pi v_h^2} \;,
\end{equation}
it can be quickly seen that for the right handed neutrino mass scales we consider and the masses of the active neutrinos, that this decay will occur rapidly during reheating, well before the electroweak phase transition suppresses the  sphaleron transitions.

%%%%%%%%%%%%%%%%%%%%%%%%%%%%%%%%%%%%%%%%%%%%%%%%%%%%%%%%%%%%%%%%%%%%%%%%%%%%%%%%%%%%%%%%%%%%%%%%%%%%
\section{Generated Baryon Asymmetry}

%After the period of driven motion there will be no further production or washout of the generated asymmetry, apart form that associated with expansion prior to the entropy production at the end of reheating. 
Using the approximate form of $ n $ derived in Eq. ($ \ref{approx_n} $), we can now obtain an equation for the lepton asymmetry parameter, assuming no further generation or washout of lepton number, apart from expansion.
The entropy density at the end of reheating is given by
\begin{equation}
s \;=\; \frac{2\pi^2}{45}g_{*} T_{rh}^3\;,
\end{equation}
where $ g_{*}\simeq 106.75 $, where no further non-negligible productions of entropy are considered. From Eq.~\eqref{Ch3eq:barnoden} we obtain the following approximate equation for the generated lepton asymmetry,
\begin{eqnarray}
\eta_L^\mathrm{reh}
\;=\; \frac{|n_{L}|}{s} 
\;=\; 0.04\;\frac{\lambda\phi_r^4}{\mu T_{rh}^3}\left(\frac{a_d}{a_{rh}}\right)^3
\;,
\end{eqnarray}
where the dilution factor from the time of driven motion to the end of reheating is given by,
\begin{equation}
\left(\frac{a_d}{a_{rh}}\right)^3 
=\; \left(\frac{\pi^2 g_{*}}{90}\right) \left(\frac{T_{rh}^4}{H_d^2 M_p^2}\right)\;.
\end{equation}

In order to successfully achieve Baryogenesis, we require that the $ B-L $ conserving  sphaleron transitions transmit the lepton asymmetry into the baryonic sector, and as such require that the reheating temperature is greater that 100 GeV. After sphaleron redistribution, the generated baryon asymmetry is given by,
\begin{equation}
\eta_B \;=\; \frac{28}{79}\,\eta_L \;\simeq\;
0.17\;\frac{\lambda \phi_{r}^4 T_{rh}}{\mu H_d^2 M_p^2} \;.
\end{equation}
This result can be simplified by utilizing the SSC,  
$H_d=\lambda \phi_{r}^2 H_i \Lambda / \mu^2 \Phi_{i}$,
\begin{eqnarray}
\eta_B \;\simeq\; 
10^{-10}\left( \frac{T_{rh}}{2\lambda \times 10^{8} ~\textrm{GeV} } \right)
\;,
\end{eqnarray}
hence we can generate the observed baryon asymmetry with reheating temperatures as low as $ 10^{5} $ GeV when we consider the lower limit on the parameter $ \lambda $ given in Eq.~\eqref{con_3}. This simple solution permits a wide range of inputs including the full allowed range of $ \lambda $. The allowed reheating temperature can be as high as that associated with the Hubble rate $ H_{d} $, assuming that the mass of $ \phi $ is much greater such that thermal production does not occur.

%-------------------------------------------------------------------%
%-------------------------------------------------------------------%
%-------------------------------------------------------------------%

%%%%%%%%%%%%%%%%%%%%%%%%%%%%%%%%%%%%%%%%%%%%%%%%%%%%%%%%%%%%%%%%%%%%%%%%%%%%%%%%%%%%%%%%%%%%%%%%%%%%
\section{Conclusion}

We have presented a model for Leptogenesis during reheating that utilizes the Ratchet
Mechanism, and is found to emulate the dynamics of a forced pendulum. This system consists of a complex scalar carrying lepton number, and an inflaton consistent
with the Starobinsky inflationary mechanism,  with the inflaton and the complex scalar interacting via a derivative coupling. The scalar lepton
potential violates L, and the violation of $ \mathcal{C} $ and $ \mathcal{CP} $ is introduced by the derivative coupling
interaction. The push out-of-equilibrium in this mechanism is provided by the reheating epoch,
which is a result of the coherent oscillation of the inflaton in its potential. In order for a non-zero
lepton number density to be produced, in the form of right handed neutrinos, driven motion must be induced in the phase $ \theta $, which is achieved when the Sweet Spot Condition is satisfied. The resultant asymmetry successfully explains the observed baryon asymmetry, with $ T_{rh}> 10^5 $ GeV permitted. This model also provides an origin for the masses of the neutrinos via the seesaw mechanism, with the right handed neutrinos having masses of the order $ 10^{11} \sim 10^{14} $ GeV. This provides a unique Leptogenesis mechanism where a high reheating temperature is not required despite consisting of a  seesaw scale close to that of the GUT scale.

%%%%%%%%%%%%%%%%%%%%%%%%%%%%%%%%%%%%%%%%%%%%%%%%%%%%%%%%%%%%%%%%%%%%%%%%%%%%%%%%%%%%%%%%%%%%%%%%%%%%
\section*{Acknowledgements}

The work of KB was supported by the JSPS Kaken-hi Grant Number JP 25800136 and 
Competitive Research Funds for Fukushima University Faculty (17RI017). 
NDB is supported in part by World Premier International Research Center Initiative (WPI), MEXT, Japan.

%%%%%%%%%%%%%%%%%%%%%%%%%%%%%%%%%%%%%%%%%%%%%%%%%%%%%%%%%%%%%%%%%%%%%%%%%%%%%%%%%%%%%%%%%%%%%%%%%%%%
%%%%%%%%%%%%%%%%%%%%%%%%%%%%%%%%%%%%%%%%%%%%%%%%%%%%%%%%%%%%%%%%%%%%%%%%%%%%%%%%%%%%%%%%%%%%%%%%%%%%

%%%%%%%%%%%%%%%%%%%%%%%%%%%%%%%%%%%%%%%%%%%%%%%%%%%%%%%%%%%%%%%%%%%%%%%%%%%%%%%%%%%%%%%%%%%%%%%%%%%%
%%%%%%%%%%%%%%%%%%%%%%%%%%%%%%%%%%%%%%%%%%%%%%%%%%%%%%%%%%%%%%%%%%%%%%%%%%%%%%%%%%%%%%%%%%%%%%%%%%%%

\end{document}